\documentstyle[12pt]{article}

\def\a{\alpha }

\begin{document}

\begin{center}
{\Large\bf Constraints on "Second Order Fixed Point" QCD
from the CCFR Data on Deep Inelastic
Neutrino-Nucleon Scattering}
\end{center}
\vskip 2cm

\begin{center}
{\bf Aleksander V. Sidorov
}
\\
{\it Bogoliubov Theoretical Laboratory\\
 Joint Institute for Nuclear Research\\
141980 Dubna, Russia\\
 E-mail: sidorov@thsun1.jinr.dubna.su}
\vskip 0.5cm
{\bf Dimiter B. Stamenov \\
{\it Institute for Nuclear Research and Nuclear Energy\\
Bulgarian Academy of Sciences\\
Boul. Tsarigradsko chaussee 72, Sofia 1784, Bulgaria\\
E-mail:stamenov@bgearn.acad.bg }}
\end{center}

\vskip 0.3cm
\begin{abstract}
The results of LO {\it Fixed point} QCD (FP-QCD) analysis of the CCFR data for
the nucleon structure function $~xF_3(x,Q^2)~$ are presented. The predictions
of FP-QCD, in which the Callan-Symanzik $~\beta~$ function admits a {\it second
order} ultraviolet zero at $~\alpha = \alpha_0~$  are in good agreement with
the data. Constraints for the possible values of the $~\beta~$ function
parameter $~b~$ regulating how fast $~\alpha_ s(Q^2)~$ tends to its
asymptotic value $~\alpha_{0}\ne 0~$ are found from the data.
The corresponding values of $~\alpha_{0}~$ are also determined.
Having in mind our recent " First order fixed point" QCD fit to the
same data we conclude that in spite of the high precision and the large
$~(x,Q^2)~$
kinematic range of the CCFR data they cannot discriminate between QCD and FP-QCD
predictions for $~xF_3(x,Q^2)~$.\\
\end{abstract}
\vskip 0.5 cm
\newpage

{\bf 1. Introduction.}
\vskip 4mm
The success of perturbative Quantum Chromodynamics (QCD) in the description
of the high energy physics of strong interactions is considerable. The QCD
predictions are in good quantitative agreement with a great number of data on
lepton-hadron and hadron-hadron processes in a large kinematic region (e.g. see
reviews \cite{Altarelli} and the references therein).
Despite of this success of QCD, we consider
that it is useful and reasonable to put the question: Do the present data fully
exclude the so-called {\it fixed point}  (FP) theory models \cite{Pol} ? \\

We remind that these models are not asymptotically free. The effective coupling
constant $~\alpha_{s}(Q^{2})~$ approaches a constant value
$~\alpha_{0}\ne 0~$ as $~Q^{2}\to \infty~$ (the so-called fixed point at
which the Callan-
Symanzik $\beta$-function $~\beta(\alpha_{0}) = 0~$). Using the assumption
that $~\alpha_{0}~$ is {\it small} one can make predictions for the physical
quantities in the high energy region, as like in QCD, and confront them
to the
experimental data. Such a test of FP theory models has been made  \cite{GR,BS}
by using the data of deep inelastic lepton-nucleon experiments started by
the SLAC-MIT group \cite{SLAC} at the end of the sixties and performed in the
seventies \cite{data70}. It was shown that

$~i$) the predictions of the FP theory models with {\it scalar} and {\it non-
colored (Abelian) vector} gluons {\it do not agree} with the data

$ii$) the data {\it cannot distinguish} between different forms of scaling
violation predicted by QCD and the so-called {\it Fixed point} QCD (FP-QCD), a
theory with {\it colored vector} gluons, in which the effective coupling
constant $~\alpha_{s}(Q^{2})~$ does
not vanish when $Q^{2}$ tends to infinity.\\

We think there are two reasons to discuss again the predictions of FP-QCD. First
of all, there is evidence from the non-perturbative lattice
calculations \cite{Pat} that the $\beta$- function in QCD vanishes
at a nonzero coupling
$~\alpha_{0}~$ that is small. (Note that the structure of the
$\beta$-function can be studied only by non-perturbative methods.) Secondly,
in the last years the accuracy and the kinematic region of deep inelastic
scattering data became large enough, which makes us hope that discrimination
between QCD and FP-QCD could be performed.\\

Recently we have analyzed the CCFR deep inelastic neutrino-nucleon scattering
data \cite{prep1} in the framework of the {\it Fixed point} QCD.
It was demonstrated \cite{SiSt} that the data for
the nucleon structure function $~xF_3(x,Q^2)~$ are in good agreement with
the LO predictions of this theory model using the assumption that the
$~\beta~$ function has a {\it first order} ultraviolet zero (fixed point)
at $~\alpha = \alpha_{0}~$ that is small.

Having in mind that up to now the structure of the fixed point theory is not
well known, it seems to us to be useful to make predictions for the
physical quantities studying the different hypotheses about the $~\beta~$
function behaviour near it fixed point $~\alpha_0~$ and confront them
to the data.\\

In this letter we present a leading order {\it Fixed point} QCD analysis of the
CCFR data \cite{prep1}, in which an expression for $~xF_3(x,Q^2)~$ based on
the assumption that the Callan-Symanzik $~\beta~$ function has a {\it second
order} ultraviolet zero at $~\alpha = \alpha_0~$ is used.
We remind that the structure function $~xF_3~$ is a pure non-singlet and the
results of the analysis are independent of the assumption on the shape of
gluons.
As in a previous analysis the method \cite{Kriv} of reconstruction of the struct
functions from their Mellin moments is used. This method is based on the Jacobi
polynomial expansion \cite{Jacobi} of the structure functions.
In \cite{KaSi} this method has
been already applied to the QCD analysis of the CCFR data.\\

{\bf 2. Method and Results of Analysis.}
\vskip 4mm
Let us start with the basic formulas needed for our analysis.

The Mellin moments of the structure function $~xF_3(x,Q^2)~$ are defined as:
\begin{equation}
M_n^{NS}(Q^2)=\int_{0}^{1}dxx^{n-2}xF_{3}(x,Q^2)~,
\label{mom}
\end{equation}
where $~n=1,2,3,4,...~.$

In the case of FP-QCD  the effective coupling constant $~\alpha_s(Q^2)~$ at
large $~Q^2~$ takes the form:
\begin{equation}
\alpha_s(Q^2) = \alpha_0 + f(Q^2)~,
\label{apre}
\end{equation}
where $~f(Q^2)\rightarrow 0~$ when $~Q^2\rightarrow\infty~.$

Let us assume that $~\alpha_0~$ is a {\it second order} ultraviolet fixed point
for the $\beta$-function, i.e.
\begin{equation}
\beta(\alpha) = -b(\alpha - \alpha_0)^2~,~~~~~~b~>~0~.
\label{beta}
\end{equation}

Then
\begin{equation}
\alpha_s(Q^2) = \alpha_0 +
{\alpha_s(Q^2_0)-\alpha_0\over 1 + b~[\alpha_s(Q^2_0)-\alpha_0]~
ln~({Q^2/Q_0^2})}~
\label{aq2}
\end{equation} \\
and we obtain for the moments of $~xF_3~$ the following leading order
expression:
\newpage
\begin{equation}
M_{n}^{NS}(Q^2)
 =M_{n}^{NS}(Q_{0}^2) \left [ \frac{Q_0^{2}}
 {Q^{2}}    \right ]^{\frac{1}{2}d^{NS}_{n}}{\cal F}_n(Q^2)~,
\label{mfp}
\end{equation}\\
where
\begin{equation}
{\cal F}_n(Q^2) = \left [ {1\over 1 + b~[\alpha_s(Q^2_0)-\alpha_0]~
ln~({Q^2/Q_0^2})} \right ]^{{d^{NS}_{n}/2b\alpha_0}}~.
\label{cf}
\end{equation}\\

In (\ref{mfp}) and (\ref{cf})
\begin{equation}
d_n^{NS} = \frac{\alpha_0}{4\pi}\gamma^{(0)NS}_n
\label{dn}
\end{equation}
and
\begin{equation}
\gamma^{(0)NS}_{n} ={8\over 3}[1 - {2\over n(n+1)} + 4\sum_{j=2}^{n}
{1\over j}]~.
\label{goa0}
\end{equation}

The $n$ dependence of $~\gamma^{(0)NS}_{n}~$ is
exactly the same as in QCD, if we assume that the Perturbative QCD expansion
for the anomalous dimensions $~\gamma^{NS}_{n}(\alpha_{s})~$ is valid in the
range of small $~\alpha_{s}~$ including the fixed point $~\alpha_0~$ too.
However, the $~Q^2~$ behaviour of the moments is different because of the
different $~Q^2~$ behaviour of the effective coupling constant
$~\alpha_s(Q^2)~$ in FP-QCD. In contrast to QCD, the Bjorken scaling for
the moments of the structure functions is broken by powers in $~Q^2~$ in
addition to the usual $~ln~Q^2~$ terms. In (\ref{cf}) and (\ref{dn})
$~\alpha_0~$ and $b$ are parameters, to be determined from the data.\\


Having in hand the moments (\ref{mfp}) and following the method
\cite{Kriv,Jacobi}, we can write
the structure
function $~xF_3~$ in the form:
\begin{equation}
xF_{3}^{N_{max}}(x,Q^2)=x^{\a}(1-x)^{\beta}\sum_{n=0}^{N_{max}}\Theta_n ^{\a ,
 \beta}
(x)\sum_{j=0}^{n}c_{j}^{(n)}{(\a ,\beta )}
M_{j+2}^{NS} \left ( Q^{2}\right ),   \\
\label{e7}
\end{equation}
where $~\Theta^{\alpha \beta}_{n}(x)~$ is a set of Jacobi polynomials and
$~c^{n}_{j}(\alpha,\beta)~$ are coefficients of the series of
$~\Theta^{\alpha,\beta}_{n}(x)~$ in powers in x:
\begin{equation}
\Theta_{n} ^{\a , \beta}(x)=
\sum_{j=0}^{n}c_{j}^{(n)}{(\a ,\beta )}x^j .
\label{e9}
\end{equation}

$N_{max},~ \alpha~$ and $~\beta~$ have to be chosen so as to
achieve the fastest convergence of the series in the R.H.S.
of Eq. (\ref{e7}) and to reconstruct $~xF_3~$ with
the accuracy required. Following the results of \cite{Kriv} we use $~\alpha =
 0.12~,
~\beta = 2.0~$ and $~N_{max} = 12~$. These numbers guarantee accuracy better
than $~10^{-3}~$.\\

Finally we have to parametrize the structure function $~xF_3~$ at some fixed
value of $~Q^2 = Q^2_{0}~$. We choose $~xF_3(x,Q^2)~$ in the form:
\begin{equation}
xF_{3}(x,Q_0^2)=Ax^{B}(1-x)^{C}~.
\label{e10}
\end{equation}

The parameters A, B and C in Eq. (\ref{e10}) and the FP-QCD
parameters $~\alpha_{0}~$ and $b$
are free parameters which are determined by the fit to the data.

In our analysis the target mass corrections \cite{tmc} are taken into
account.
To avoid the influence of higher--twist effects we have used only the
experimental points in the plane $~(x,Q^2)~$
with $~10 < Q^2\leq 501~(GeV/c)^2~$. This cut corresponds to the following
$~x~$ range:$~0.015\leq x \leq 0.65~$.\\

The results of the fit are presented in Table 1. In all fits only statistical
errors are taken into account.\\

\vskip 0.2 cm
\begin{center}
\begin{tabular}{|c|c|c|c|c|c|c|} \hline
$b$&$\chi^2_{d.f.}$& $\alpha_0$ &$\alpha_s(M_z^2)_{FP-QCD}$  \\ \hline
 0.9 & 83.5/61&~0.029~$\pm$~0.014~&~0.123~$\pm$~0.022~\\
 1.0 & 83.3/61&~0.040~$\pm$~0.014~&~0.126~$\pm$~0.021~\\
 1.1 & 83.1/61&~0.048~$\pm$~0.014~&~0.128~$\pm$~0.021~\\
 1.3 & 82.9/61&~0.063~$\pm$~0.013~&~0.133~$\pm$~0.019~\\
 1.5 & 82.6/61&~0.074~$\pm$~0.013~&~0.136~$\pm$~0.019~\\ \hline
\end{tabular}
\vskip 4mm
\begin{tabular}{cl}
{\bf Table 1.}& The results of the LO FP-QCD fit
to the CCFR $~xF_3~$ data.\\
&$\chi^2_{d.f.}$ is the
$\chi^2$-parameter normalized to the degree of freedom $d.f.$.
\end{tabular}
\end{center}
\vskip 0.8 cm

Summarizing the results in the Table one can conclude:

1. The values of $~\chi^2_{d.f.}~$  are practically the same as in the case
of a first order ultraviolet fixed point for the $~\beta~$ function \cite
{SiSt}. They are slightly smaller
than those obtained in the LO QCD analysis \cite{KaSi} of
the CCFR data by the same method and indicate a {\it good description}
of the data.

2. It is seen from the Table that $~\alpha_0~$ increases with
increasing $~b~$.
The values of $~b~$, for which the asymptotic coupling $~\alpha_0~$ is
acceptable, are found to range in the following interval:
\begin{equation}
0 < b \leq 1.5~.
\label{ib}
\end{equation}

For the values of $~b > 1.5~$ the corresponding theoretical values
for $~\alpha(M^2_z)~$ obtained by the Eq.(\ref {aq2}) are bigger than
$~\alpha_ s(M^2_z)~$ determined from the LEP experiments \cite {Alt}:
\begin{equation}
\alpha_ s( M^2_z) = 0.125~ \pm~ 0.005~.
\label {Mz}
\end{equation}

For the values of $b$ smaller than 0.9 $~\alpha_0~$ can not be determined
from CCFR data. The errors in $~\alpha_0~$ exceed the mean values of this
parameter.

3. The accuracy of determination of $~\alpha_0~$ is not good enough.
The accuracy increases with increasing $b$.

4. The values of $\alpha_0 = 0.029,~0.040,~ 0.048~$ corresponding to
$~b = 0.9,~1.0,~1.1~$ are preferred to the
other values of $~\alpha_0~$ determined from the data.

5. The values of $~\alpha_ s(M^2_z)~$ corresponding to the values of $~b~$
from the range (\ref{ib}) are in agreement within one standard deviation
with $~\alpha_ s(M^2_z)~$ determined from the LEP experiments.

6. The values of the parameters A, B and C are in agreement with the
results of \cite {SiSt} and \cite{KaSi}. For illustration we present here
the values of these parameters at $~Q^2_0~=~3~(GeV/c)^2~$ for $~b = 1.0~:$
$$ A = 7.07~\pm~0.20~,~~~~~B = 0.865~\pm~0.013~,~~~~C = 3.43~\pm~0.004~.$$

They are found to be independent of $~b~$ and
$~\alpha_0~$. We have found also that multiplying the R.H.S. of (\ref{e10}) by
term $~(1+ \gamma x)~$ one can not improve the fit.\\

{\bf Summary.}
\vskip  4mm
The  CCFR  deep inelastic neutrino-nucleon scattering data have been
analyzed in the
framework of the {\it Fixed point} QCD. It has been demonstrated that the
data for
the nucleon structure function $~xF_3(x,Q^2)~$ are in good agreement with
the LO predictions of this quantum field theory model using the assumption
that $~\alpha_{0}~$
is a {\it second order ultraviolet fixed point} of the $~\beta~$ function and
$~\alpha_{0}~$ is {\it small}.  Some constraints on the behaviour of the
$~\beta~$ function near $~\alpha_{0}~$ have been found from  the data.
The values for $~\alpha_{0}~$
$$0.029~ \leq~\alpha_ 0~ \leq~0.048~,$$
corresponding to the $~\beta~$ function parameter $~b~$ in the range
$$0.9~\leq~b~\leq~1.1~ $$
are preferred to the other ones determined from the data.  \\

In conclusion, taking into account also our previous results \cite {SiSt}
in the case of $~\beta~$ function with a {\it first order} fixed point,
we find that the CCFR data, the most precise data on deep
inelastic scattering at present, {\it do not eliminate} the FP-QCD and
therefore other tests have to be made in order to distinguish between QCD and
FP-QCD.\\

After completion of this work we learned about the paper
\cite{pp} where the measured inclusive cross section in $~p\bar{p}~$
collisions at $~\sqrt {s}=1.8~TeV~$ has been presented. It has
been shown that the cross section for jets with $~E_t > 200~GeV~$
is significantly higher than predictions based on perturbative QCD.

Having in mind that at such large energies $~\alpha_s(Q^2)~$ in
FP-QCD runs slower than its perturbatively predicted rate we hope
this fact could help to explain the deviation from the
perturbative QCD predictions mentioned above.\\

\vskip 4mm
{{ \bf Acknowledgments}}
\vskip 3mm
We are grateful to A.L. Kataev and C.V. Mihaylov for useful discussions
and remarks. One of us (D.S) would like to thank also I.T. Todorov for
calling his attention to the papers of A. Patrascioiu and E. Seiler, in
which the problem on vanishing of the $~\beta~$ function in QCD at
nonzero coupling is discussed.\\

This research was partly supported by
the Russian Fond for Fundamental Research Grant N 96-02-17435-a
and by the Bulgarian Science Foundation under Contract \mbox{Ph 510.}\\

\end{document}